\documentclass[aps,pra,twocolumn,superscriptaddress]{revtex4}

\usepackage{graphicx}
\usepackage{epsfig}
\usepackage{amsmath}
\usepackage{bm}
\usepackage{multirow}
\usepackage{mathrsfs}

\begin{document}

\title{ Extended Bose-Hubbard model in a shaken optical lattice }
\author{Jiao Miao}
\affiliation{Institute for Advanced Study, Tsinghua University, Beijing, 100084, China}

\begin{abstract}
We study an extended Bose-Hubbard model with next-nearest-neighbor (NNN) hopping in a shaken optical lattice. We show how mean-field phase diagram evolves with the change of NNN hopping amplitude $t_{2}$, which can be easily tuned via shaking amplitude. As $t_{2}$ increases, a $Z_{2}$-symmetry-breaking superfluid ($Z_{2}$SF) phase emerges at the bottom of the Mott lobs. The tricritical points between normal superfluid, $Z_{2}$SF, and Mott insulator (MI) phases are identified. We further demonstrate the tricritical point can be tuned to the tip of the Mott lobe, in which case a new critical behavior has been predicted. Within random-phase approximation, excitation spectra in the three phases are obtained, which indicate how the phase transitions occur.
\end{abstract}

\maketitle

\section{introduction}
Ultracold atoms condensed in periodically shaken optical lattices have shown novel properties. Two kinds of lattice shaking techniques have been developed. One is the off-resonant lattice shaking, in which the shaking frequency is tuned to be very large compared to the band gap and width. The hopping parameters and the interparticle interactions can be tuned by lattice shaking, which could result in synthetic gauge fields \cite {Hemmerich,Sengstock1,Sengstock2,Sengstock3}, an effective attractive Fermi-Hubbard model \cite{Tsuji}, or topologically nontrivial phases \cite{haldane1,haldane2}.

The other is the near-resonant lattice shaking, in which the shaking frequency is tuned to be a little larger than the gap of two energy bands. In this case different Bloch bands are hybridized, which will dramatically modifies the single-particle  dispersion and leads to interesting phenomena. In a shaken one-dimensional optical lattice, a $Z_{2}$-symmetry-breaking superfluid ($Z_{2}$SF) phase has been observed \cite{chin}, and an effective field theory has been constructed to study the normal superfluid-(NSF-)$Z_{2}$SF-Mott insulator (MI) phase transition \cite{sf ising}. And the effective theory predicted a new critical behavior nearby the tricritical point with particle-hole symmetry in three dimensions \cite{sf ising}. Algebraical orders \cite{algebraic order1, miao} and topological nontrivial phases \cite{topo} are predicted in shaken higher-dimensional optical lattices.

The Bose-Hubbard (BH) model, which consists of nearest-neighbor (NN) hopping and on-site interaction, is used to study a MI-superfluid transition \cite{bhm}. The model is a good approximation in the tight-binding limit and has been realized in an optical lattice \cite{sf-mi exp}. Considerable efforts have been dedicated to extend the model by adding terms, such as next-nearest-neighbor (NNN) hopping \cite{t1t2, t1t2_2}, nearest-neighbor interaction \cite{nn int}, dipolar interaction \cite{dipolar int}, interaction-induced hopping term \cite{int hopping}, spin structure \cite{spin}, or disorder \cite{disorder1,bhm}. The NN and NNN hopping parameters can be renormalized in a different way  by off-resonant lattice shaking, hence the ratio between them can be tuned \cite{t1t2_2}.

In this paper, we show an extended Bose-Hubbard (EBH) model with NNN hopping can be easily realized by shaking optical lattices resonantly. Within mean-field theory, we find  NSF, $Z_{2}$SF and MI phases. We further show $Z_{2}$SF phase emerges at the bottom of the Mott lobes for nonvanishing NNN hopping amplitude. In three dimensions, a new critical exponent of superfluid transition is predicted near the tricritical point with particle-hole symmetry \cite{sf ising}. Nevertheless, the analysis is based on a constructed effective theory, and the existence of the particle-hole-symmetric tricritical point is in doubt. Here within the microscopic EBH model, we demonstrate the tricritical point always exists and can be tuned to the tip of a Mott lobe. This makes previous work \cite{sf ising} more reliable. In the end, we calculate excitation spectra in the MI and superfluid phases in the random-phase approximation. We find gapless superfluid excitation has a quadratic dispersion near the condensate momentum at the NSF-$Z_{2}$SF transition boundary. We also demonstrate in the $Z_2$SF phase that the excitation spectrum has a roton structure in the strong-coupling limit.

\section{mean-field phase diagram}
Let us consider a Chicago-type experiment \cite{chin}. Two counterpropagating laser beams are time-periodically modulated. The Hamiltonian \cite{chin}
reads
\begin{align}
H(t) = \frac{ p_x^2 }{ 2 m } + V \cos^2 ( k_r x + \frac{ \theta(t) }{2} ),
\end{align}
where $\hbar k_r$ is the photon momentum, $\theta(t) = f \cos ( \omega_0 t )$,
$f$ and $\omega_0$ are shaking amplitude and frequency, respectively, and $\Delta \equiv f / ( 2 k_r )$ is the maximum displacement of the lattice.

By performing a transformation, $x \rightarrow x - \Delta \cos ( \omega_0 t ) $, in the comoving frame the Hamiltonian reads
\begin{equation}  \label{Ht}
H (t) = \frac{ p_x^2 }{ 2 m } + V \cos^2 ( k_r x ) - \frac{ A_x(t) p_x}{m},
\end{equation}
where $A_x(t)= m \omega_0 \Delta \sin ( \omega_0 t ) $. An ac electric field $E_x = - m \omega_0^2 \Delta \cos (\omega_0 t)$ is effectively imposed to bosons condensed in the unshaken lattice.
The first two terms, representing the unshaken lattice, give a static band structure $\epsilon_{\lambda}( k_x ) $ with Bloch state $\Psi_{\lambda,k_x} (x) $. We choose the Bloch state as a basis in our following analysis. 

\begin{figure}[htb]
\centering
\includegraphics[width=3.2in]{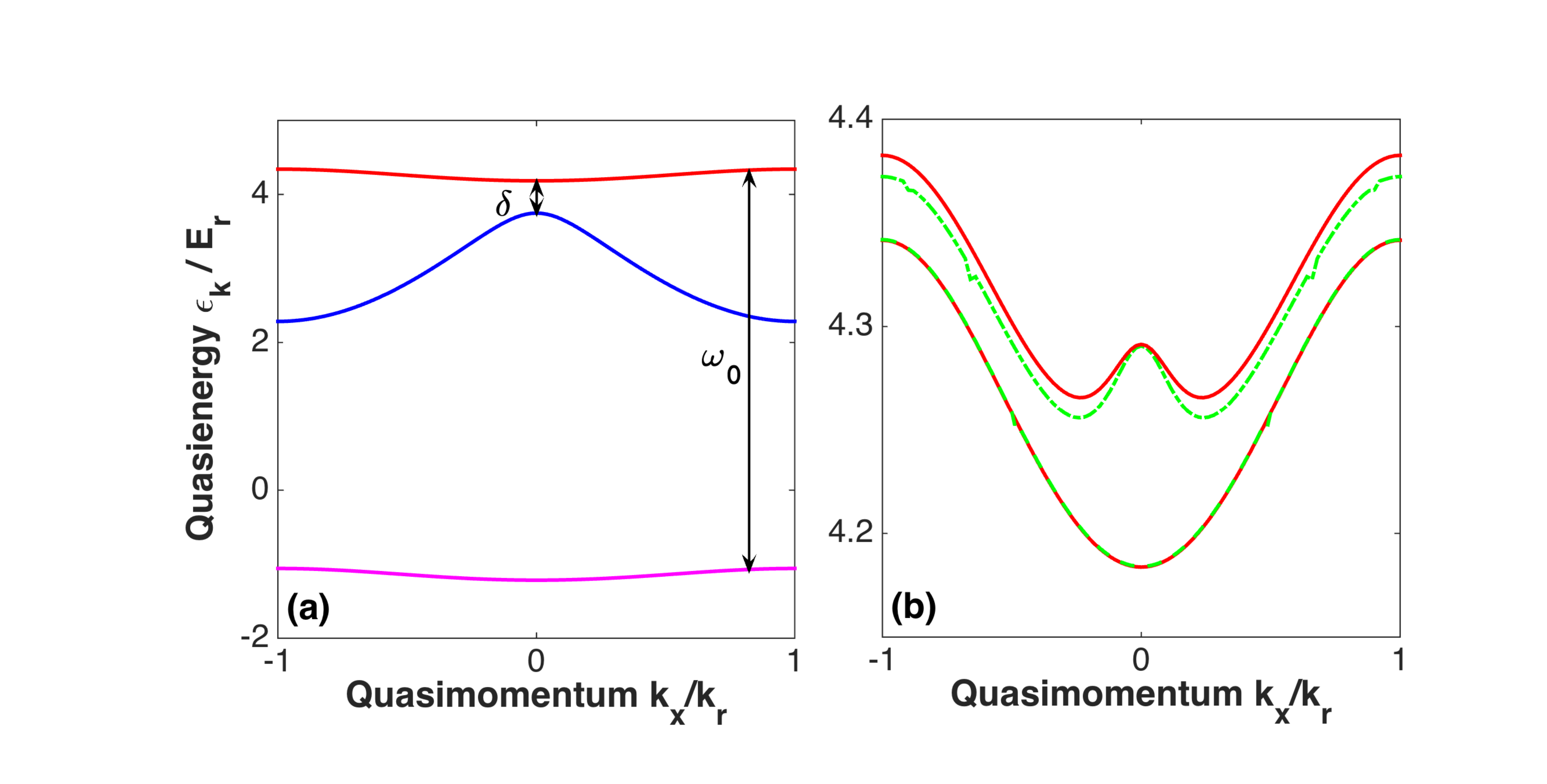}
\caption{Band structure with lattice depth $V= 7 E_r $ and detuning $\delta = 0.44 E_r$, where $E_r=(\hbar k_x)^2 / (2m)$ denotes photon recoil energy. (a) Band structure before shaking. The magenta and blue lines denote $s$ and $p$ bands, respectively. The red line denotes the dressed $s$ band with a detuning $\delta$.  (b) Quasienergy dispersion of the upper hybridized band. The red and green lines are calculated using 2 and 21 bands, respectively.  The upper and lower lines with the same color denotes the dispersion for shaking amplitude $f=0.2$ and $ f=0$, respectively.}
\label{dispersion}
\end{figure}

In the experiment \cite{chin}, shaking frequency is tuned to make $s$ and $p$ bands near-resonant. So we will use two-band and rotating-wave approximations in the following analysis. A quasienergy spectrum obtained by numerically diagonalizing Floquet operator $ \mathcal{T} \textrm{exp} \{ - \frac{i}{\hbar} \int_0^{T} H(t) \} $ for the lowest 21 bands \cite{chin} is shown in Fig. \ref{dispersion} (b), where $ \mathcal{T} $ denotes time ordering and $T = 2 \pi / \omega_0$ is time period. Fig. \ref{dispersion} (b) indicates the approximations are very good. 

The Hamiltonian in the tight-binding form reads
\begin{align}
H(t)=\sum_{k_x} \left( \Psi^{\dagger}_{p,k_x}, \Psi^{\dagger}_{s, k_x}
\right) H_{k_x}(t) \left(
\begin{array}{c}
\Psi_{p, k_x} \\
\Psi_{s, k_x}
\end{array}
\right),
\end{align}
where
\begin{align}
H_{k_x}(t) & = \left(
\begin{array}{cc}
\epsilon_p(k_x) & 0 \\
0 & \epsilon_s(k_x)
\end{array}
\right) + \sin(\omega_0 t)
\notag \\
& \quad  \times \left(
\begin{array}{cc} 
- 4 h_p \sin ( k_x d ) & -2 i \Omega_{k_x} \sin (\omega_0 t) \\
2 i \Omega_{k_x} \sin (\omega_0 t)  &  -4 h_s \sin ( k_x d ) 
\end{array}
\right),
\\
\Omega_{k_x} & = h_{sp} + h_{sp1} \cos ( k_x d ), \\
h_{sp} & = -\frac{\omega_0 \Delta}{2} \langle w_p (x) | i p_x | w_s (x)
\rangle, \\
h_{sp} & = -\frac{\omega_0 \Delta}{2} \langle w_p (x) | i p_x | w_s (x-d)
\rangle,
\end{align}
$\Psi^{\dagger}_{\lambda, \bm{k}}$ is creation operator of a boson in
the $\lambda$-band with quasimomentum $k_x$, $d = \pi / k_r$ is the lattice constant, $w_{\lambda}$ is the Wannier function for the $\lambda$ band, $\lambda$ denotes $s$ or $p$, and  $\langle \cdots | \cdots | \cdots \rangle$ denotes a real-space integral $\int dx \cdots$.

In the rotating-wave approximation, the effective Hamiltonian reads
\begin{align}
H_{k_{x}}& = \mathscr{U}^{\dagger }(t)[H_{k_{x}}(t)-i\partial _{t}]
\mathscr{U}(t)  \notag  \label{Heff} \\
& \approx \left(
\begin{array}{cc}
\epsilon _{p}(k_{x}) & \Omega _{k_{x}} \\
\Omega _{k_{x}} & \epsilon _{s}(k_{x})+ \hbar \omega_0
\end{array}
\right),
\end{align}
where
\begin{equation*}
\mathscr{U}(t)=\left(
\begin{array}{cc}
1 & 0 \\
0 & e^{i\omega_0 t}
\end{array}
\right).
\end{equation*}%
Here we neglect the fast-rotating terms. Before lattice shaking, the $s$ band is
decoupled with the $p$ band due to inversion symmetry (IS). We notice lattice shaking effectively breaks IS, causing the coupling between $s$ and $p$ bands. Lattice shaking plays the same role with the electric field applied in the orbital Rashba effect \cite{rashba}.

The quasienergy spectrum calculated by diagonalizing the effective Hamiltonian in Eq. (\ref{Heff}) is shown in Fig. \ref{dispersion}. Before shaking, the dressed $s$ band has a perfect $\cos ( k_x d )$-type dispersion, and therefore NNN hopping can be neglected. Lattice shaking changes the dressed $s$ band into a hybridized band, in which bosons will stay when turning on shaking adiabatically. As shaking amplitude increases, the hybridized band dispersion deviates from the $\cos( k_x d )$ form. So an extra $\cos (2 k_x d )$ term needs to be considered.

Assuming the ground state is $\Psi_{k_c}(x)$ with quasimomentum $k_c$, which
breaks $Z_2$ symmetry spontaneously for nonvanishing $k_c$ \cite{sf ising}, time-average
interaction energy reads
\begin{align}
\epsilon_{int}(k_c) = \frac{1}{T} \int_0^{T} dt \, g \int dx | \Psi_{k_c} (x)
|^4,
\end{align}
where $g$ is the repulsive interaction strength.

\begin{figure}[htb]
\includegraphics[width=2.5in]
{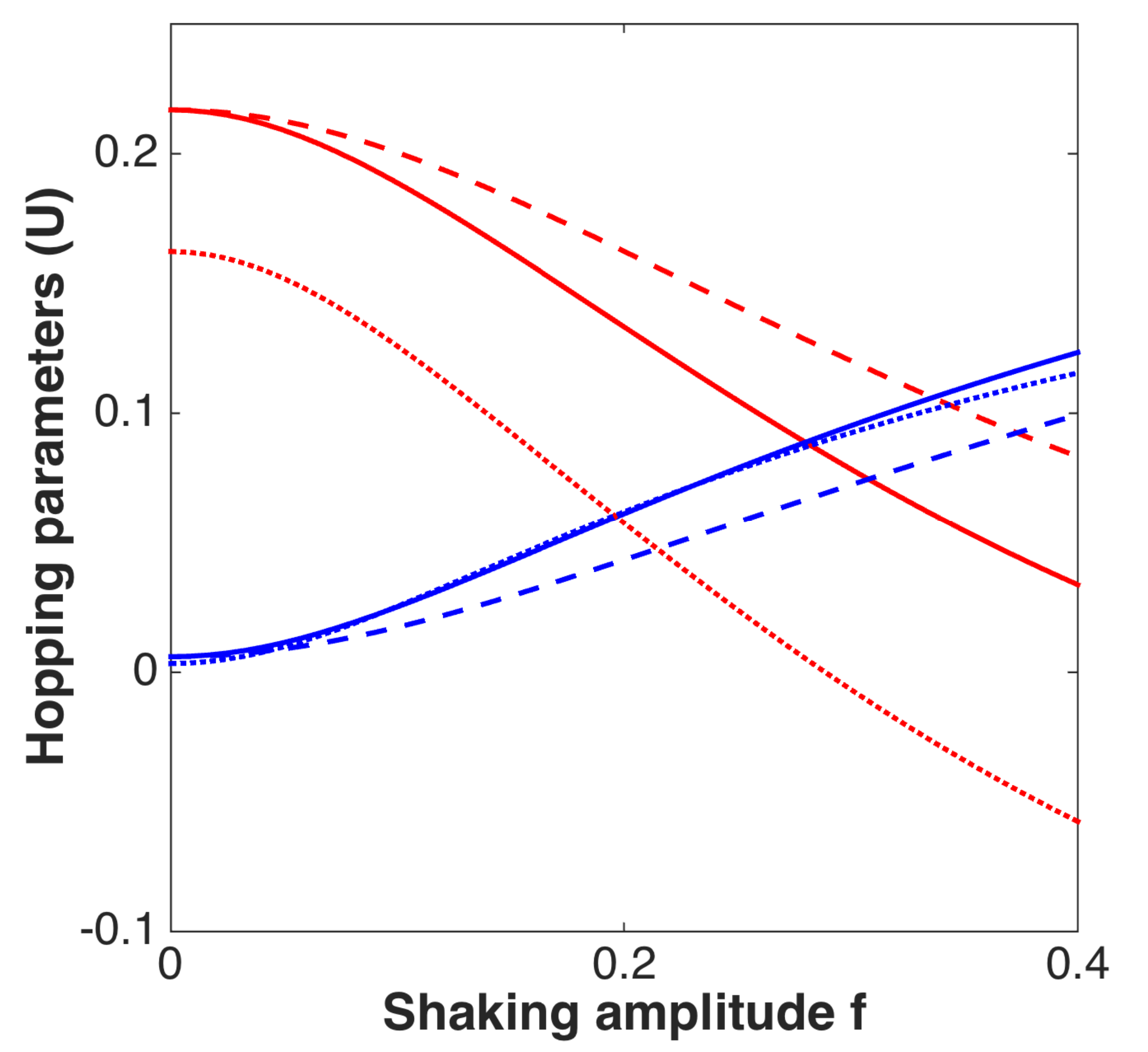}
\caption{Lattice shaking induced NN and NNN hopping. Parameters $(V/E_r, \delta/E_r)$ for dotted, solid, and dashed lines are $(8, 0.25)$, $(7, 0.25)$, and $(7, 0.45)$, respectively. Interaction energy is $g n = 0.1 E_r$, where $n$ denotes particle density. The red and blue lines denote $t_1/U$ and $t_2/U$, respectively, where $U$ denotes on-site interaction energy.}
\label{t_f}
\end{figure}

We project Hilbert space into the upper hybridized band. In the tight-binding limit, the next-next-nearest-neighbor hopping strength  is smaller than the NNN hopping strength and will be neglected without affecting the following qualitative results. The off-site interaction energy is much smaller than the on-site interaction energy and will also be neglected. So we will study an EBH Hamiltonian
\begin{align} \label{h_ebh}
H_{EBH} &= - t_1 \sum_{<i,j>} a^{\dagger}_i a_j + t_2 \sum_{<<i,j>>}
a^{\dagger}_i a_j - \mu \sum_i n_i  \notag \\
& \quad + \frac{U}{2} \sum_i n_i ( n_i -1 ),
\end{align}
where $a_i$ is the boson operator annihilating boson at site $i$, $n_i = a_i^{\dagger} a_i $ is boson number operator, $\mu$ is the chemical potential, $U$ is the on-site interaction, and the summations for the first and second terms are over NN and NNN sites, respectively. 

 Within standard mean-field theory, the EBH Hamiltonian in Eq. (\ref{h_ebh}) can be rewritten as
\begin{equation}  \label{H_ebh}
H_{EBH} =  H_{MF} - t_1 \sum_{<i,j>}  \tilde{a}^{\dagger}_i \tilde{a}_j + t_2 \sum_{<<i,j>>} \tilde{a}^{\dagger}_i \tilde{a}_j,
\end{equation}
where
\begin{align}  
H_{MF} & \equiv \sum_i H_{MF}^i,     \label{H_mf}
\\
H_{MF}^i & = 2 \tilde{t} \psi^2 - 2 \tilde{t} ( \psi_i  a^{\dagger}_i + \psi^{\ast}_i a_i ) - \mu n_i  \notag \\
& \quad + \frac{U}{2} n_i ( n_i -1 ),     \label{H_mf_i}
\end{align}
$\tilde{a}_i = a_i - \psi_i $ represents fluctuation, and $\tilde{t} = t_1 \cos(k_{xc} d) - t_2 \cos(2 k_{xc} d)$. The order parameter $ \psi_i \equiv \langle a_i \rangle = e^{i k_{x c} x_i} \psi $ is site-dependent, where $\langle \cdots \rangle$ denotes expectation value in the mean-field ground state, $k_{x c}$ is the condensate momentum in the $x$ direction, $x_i$ is the $x$ coordinate of the $i$th lattice site, and $\psi$ is positive and uniform. The mean-field Hamiltonian $H_{MF}$ breaks U(1)$\times Z_2$ symmetry when $\psi$ and $k_{xc}$ are nonvanishing. There are three possible phases: (1) MI phase with  $\psi=0$, (2) NSF phase with $\psi \neq 0$ and $k_{x c}=0$, and (3) $Z_2$SF phase with $\psi \neq 0$ and $k_{x c} \neq 0$.

By minimizing the single-particle dispersion $\epsilon(k_x) = - 2 t_1 \cos(k_x d) + 2 t_2 \cos (2 k_x d)$ with respect to $k_x$, in superfluid phase ($\psi \neq 0 $), $k_{x c}$ has the value of
\begin{equation}  \label{phi}
k_{x c} = \left \{
\begin{array}{ll}
0, & t_1 \ge 4 | t_2 | \\
\frac{1}{d} \text{arccos} \frac{t_1}{4 t_2}, & | t_1 | < 4 | t_2 | \\
\frac{\pi}{d}, & t_1 \le - 4 | t_2 |
\end{array}
\right. .
\end{equation}
The critical shaking amplitude $f_c$ of the NSF-$Z_2$SF transition is determined by the condition 
\begin{equation}\label{condition}
t_1 = 4 | t_2 |.
\end{equation}

Fig. \ref{t_f} shows that as shaking amplitude increases, NN hopping parameter $t_1$ decreases, while NNN hopping parameter $t_2$ increases. Here the on-site interaction energy $U$ is almost a constant for small shaking amplitude $f$.
For a fixed detuning, as lattice depth $V$ decreases, initial $t_1^0$ before shaking increases. Initial $t_2^0$ before shaking is almost vanishing. When the detuning $\delta$ is fixed and $f$ is small, the degree of the hybridization and the slope of the line $t_{1,2}$-$f$ at any $f$ are nearly the same for different $V$. So an increased $f_c$ is needed for a decreased $V$ to meet the transition condition in Eq. (\ref{condition}). For a fixed $V$, as $\delta$ increases, $t_{1,2}^0$ remains the same, the change of $t_{1,2}$ with respect to $f$ gets slower, and hence $f_c$ increases. 

We numerically calculate the order parameter $\psi_i$ in Eq. (\ref{H_mf_i}) by the standard self-consistent approach and find the MI-superfluid transition is a second-order transition. So we can use the Landau theory \cite{landau} of phase transitions. By using perturbation theory near the MI phase boundary for small $\psi$, we obtain the mean-field ground state energy
\begin{align}
\frac{E(\psi)}{M} & = - \mu n + \frac{U}{2} n(n-1) + 2 \tilde{t} \left[ 1 - 2 \tilde{t} \chi (\mu,n) \right] \psi^2 + O( \psi^4 ),
\end{align}
where $M$ is the number of lattice sites and $\chi( \mu / U, n )= \frac{n+1}{ U n - \mu } + \frac{n}{ \mu - U (n-1) }$. So the MI-superfluid transition boundary is given by
\begin{align}  \label{MI boundary}
1 - 2 \tilde{t} \chi (\mu / U, n) = 0.
\end{align}
And the boundary condition can be rewritten as
\begin{align}  \label{MI boundary1}
\frac{ \mu_{\pm} }{ U } & = - \frac{1}{2} + n - \frac{ \tilde{t} }{U} \pm \frac{1}{2} \sqrt{ 1 - 4 (1+2n) \frac{ \tilde{t} }{U} + 4 ( \frac{ \tilde{t} }{U} )^2 }.
\end{align}

\begin{figure}[htb]
\centering
\includegraphics[width=3.5in]{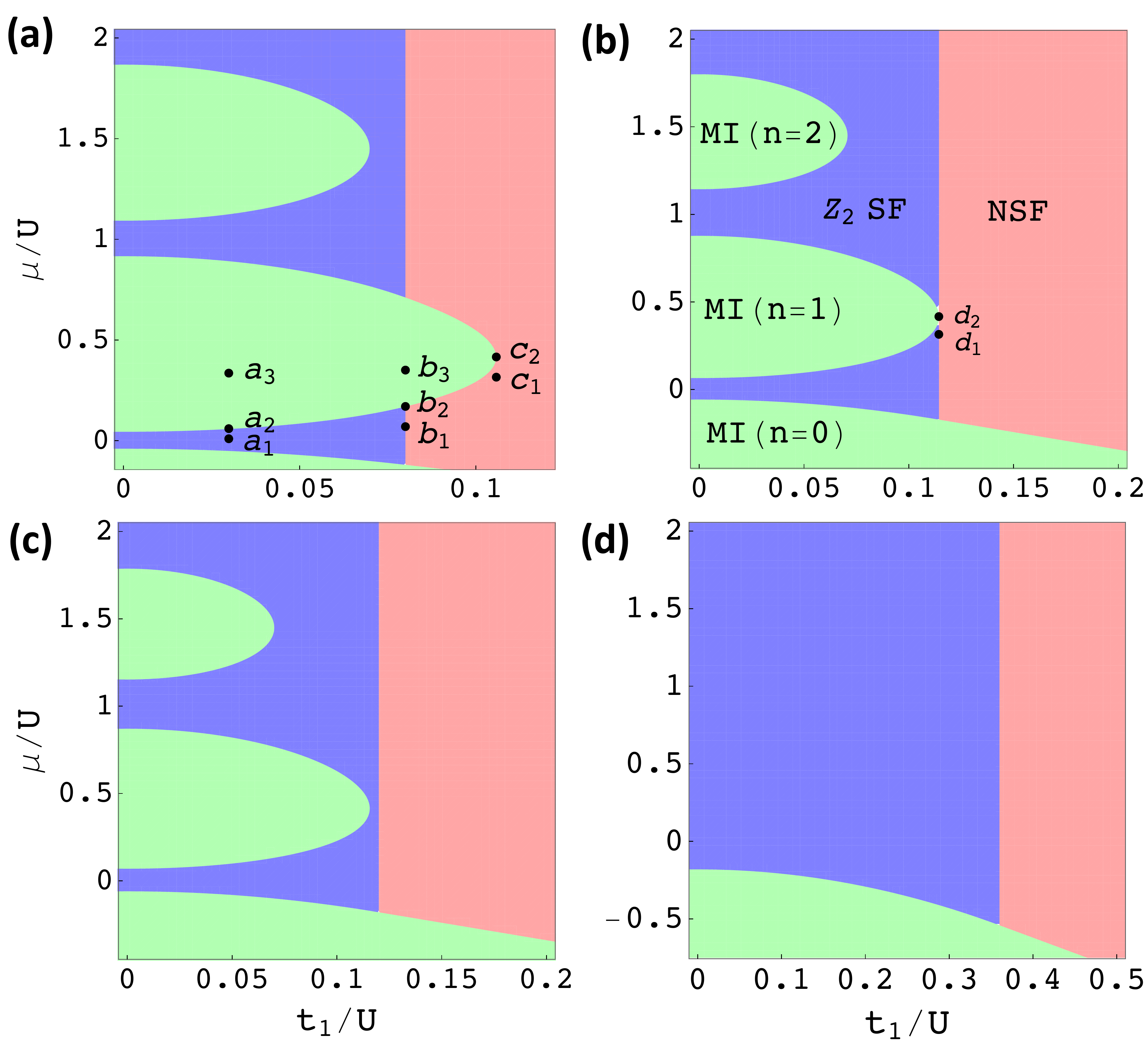}
\caption{Phase diagram for different $t_2$. $t_2$ increases from (a) to (d). The green, red, and blue regions denote the regions of MI, NSF, and $Z_2$SF phases, respectively. }
\label{phase_diagram}
\end{figure}

The mean-field phase diagram for a fixed $t_2$ is shown in Fig. \ref{phase_diagram}. As $t_2$ increases from zero, $Z_2$SF phase begins to appear at the bottom of Mott lobes near integer values of $\mu /  U$. Tricritical points lie on sides of Mott lobes. For a fixed filling number $n$, the tip of the Mott lobe lies at chemical potential $ (\mu / U )_c = \sqrt{ n^2 + n } - 1 $, which is the same as that in the standard BHM. The $Z_2$SF region grows and the Mott lobe gets thinner and longer because of competition between NN and NNN hopping. For a critical NNN hopping amplitude $(t_2 / U)_c  = 1 / [ 6 U \chi( (\mu/U)_c, n) ] $, the tricritical point coincides with the tip of the Mott lobe. When $t_2$ continues to increase, the Mott lobe gets first longer and then shorter and finally vanishes. Here the microscopic theory supports the existence of the tricritical point in Ref. \cite{sf ising}. The Mott lobe is the longest for $( t_2 / U )_l = 1 / [ 4 U \chi( (\mu /U)_c, n) ]$, which is larger than $t_2^c$. The critical behavior near a particle-hole-symmetric tricritical point is usually different from the mean-field results and attracts a lot of interest. In three dimensions, a $O(2)$ rotor universality class \cite{miao} and a new universality class \cite{sf ising} have been predicted. For a given $n$, one can tune the parameters to $[ ( \mu / U )_c , (t_1 / U)_c , (t_2 / U)_c ]$, where $(t_1 / U)_c = 2 / [ 3 U \chi( (\mu/U)_c,n) ]$, to make the tricritical point meet the tip of the Mott lobe.

The Mott lobes have varying shapes and fixed chemical potentials for their tips for different NNN hopping in the mean-field level. A beyond-mean-field theory using a $U(1)$ quantum rotor approach has predicted the same result of bosons with NNN hopping in a two-dimensional square lattice \cite{t1t2}. The approach only describes the $U(1)$ symmetry-breaking NSF-MI phase transition.

\begin{figure}[htb]
\includegraphics[width=3in]
{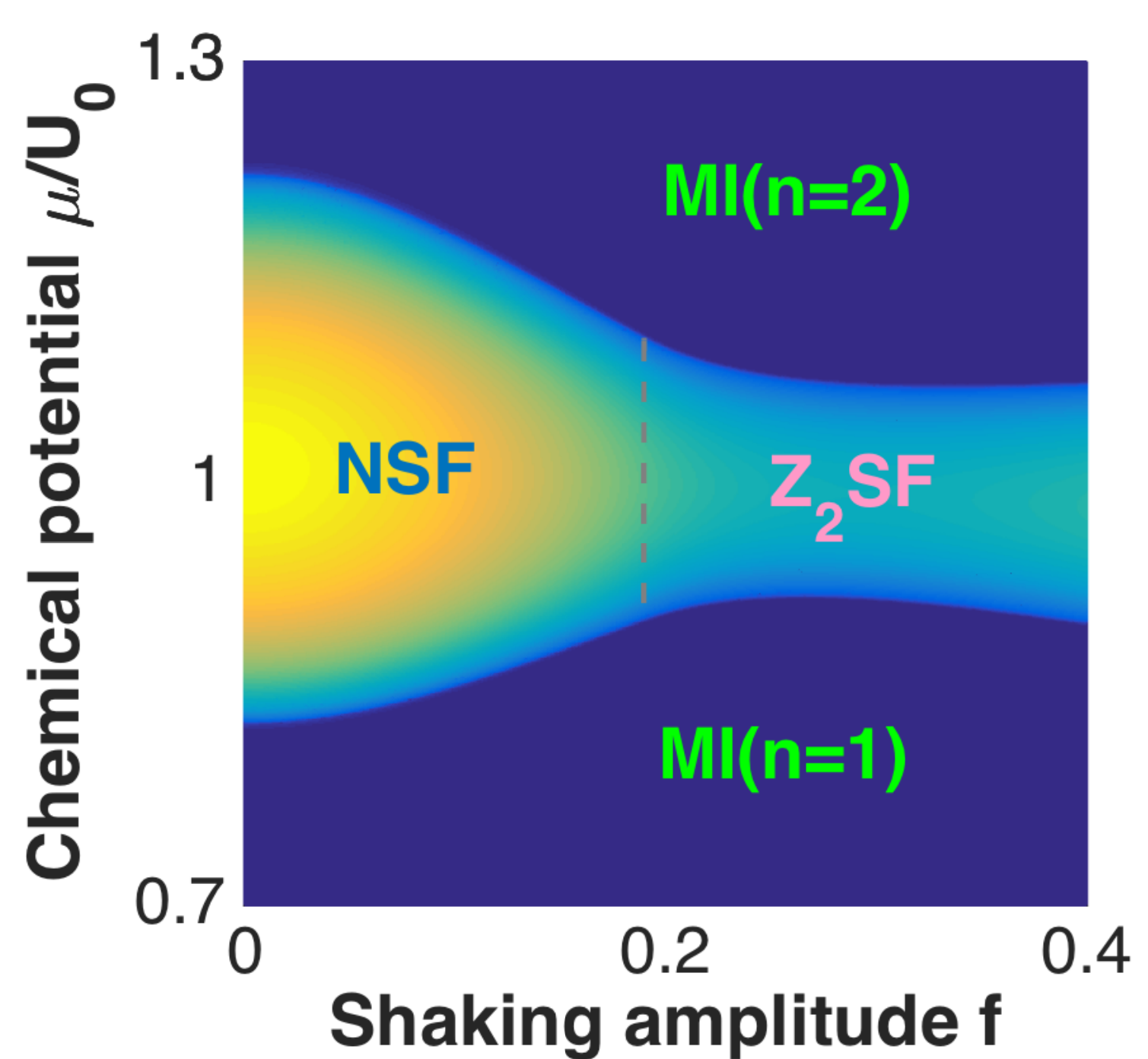}
\caption{Mean-field phase diagram with lattice depth $V=7 E_r$ and shaking
frequency $\omega_0=5.4 E_r / \hbar$. $U_0$ is the on-site interaction energy
before shaking.}
\label{f_diagram}
\end{figure}

\begin{figure}[htb]
\includegraphics[width=3.3in]
{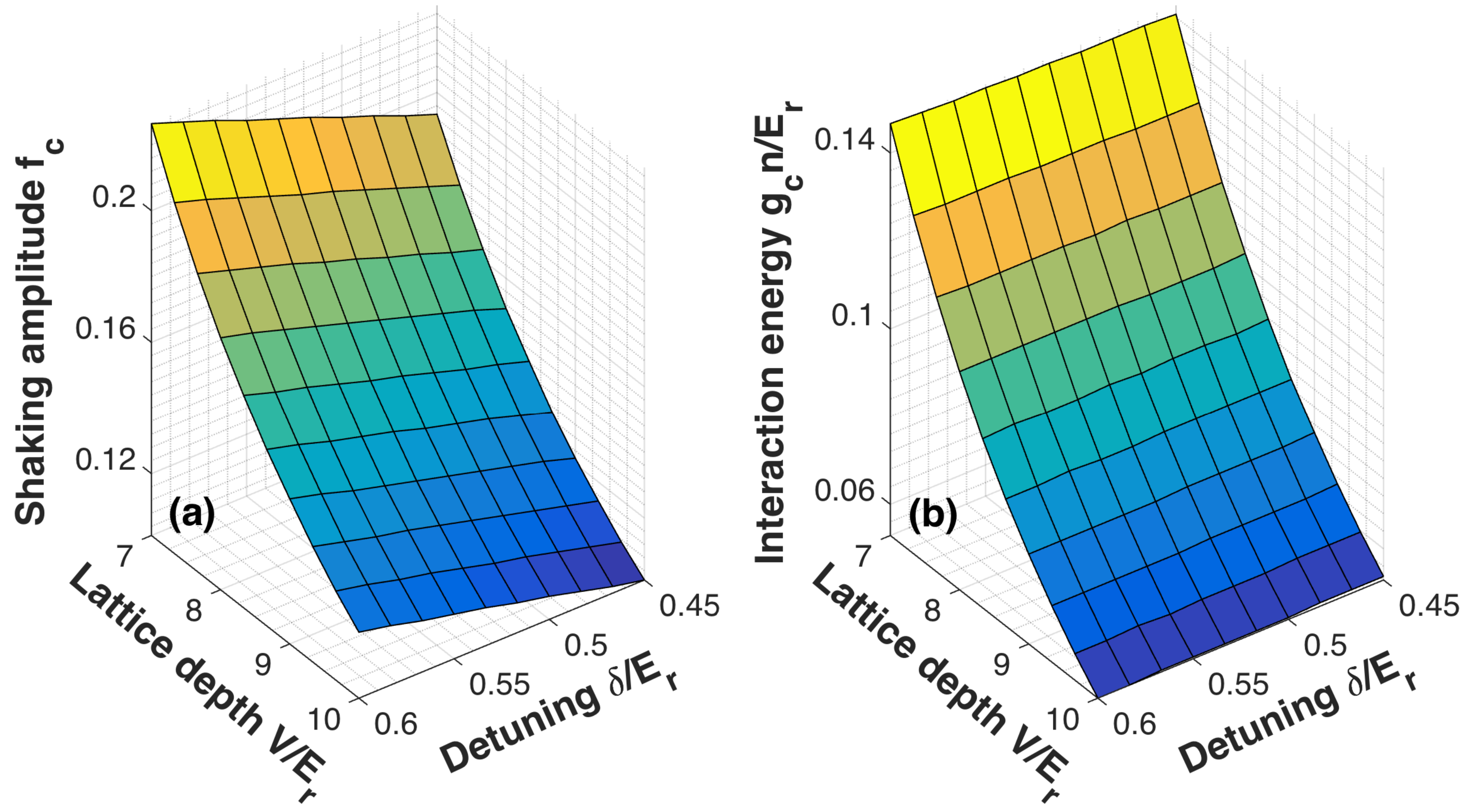}
\caption{Parameters regime for a tricritical point at the tip of the $n=1$ Mott lobe.}
\label{tricritical_regime}
\end{figure}

Fig. \ref{f_diagram} shows the phase diagram expressed in $f$ and $\mu$ terms. When the shaking amplitude $f$ increases, $t_1$ decreases and $t_2$ increases. So an initial NSF phase can turn into a $Z_2$SF phase, and the tricritical point can be tuned onto the tip of a Mott lobe.  The parameter regime, where tricritical point at the tip of the $n=1$ Mott lobe lies, is shown in Fig. \ref{tricritical_regime}. Figure \ref{tricritical_regime} (a) shows the relationship between critical shaking amplitude $f_c$ and lattice depth $V$ (detuning $\delta$) as discussed before. We know critical $ ( t_1 / U )_c$ is a constant in the parameter regime. From previous analysis we also know $ t_1 $ at the NSF-$Z_2$SF transition boundary increases as $V$ decreases and does not change much for small $\delta$. And $U$ is proportional to interaction strength $g$ and changes little with $V$ and $\delta$. So the critical interaction strength $g_c$ in the parameter regime increases as $V$ decreases and changes little with $\delta$ under the near-resonant condition, as shown in Fig. \ref{tricritical_regime} (b). When a  shaken one-dimensional lattice system is tuned to this tricritical point, one can measure the critical exponent via the $\mathit{in}$ $\mathit{situ}$ technique \cite{in situ} and a new universality class is expected \cite{sf ising}.

\section{collective excitations}
In this section, we study collective excitations at zero temperature. Following the standard-basis operator approach \cite{standard basis, rpa}, we choose eigenstates $\{ | i \alpha \rangle \}$ of the single-site mean-field Hamiltonian $H_{MF}^i$ in Eq. (\ref{H_mf_i}) as a basis, and the EBH Hamiltonian $H_{EBH}$ in Eq. (\ref{H_ebh}) can be rewritten as
\begin{align}
H_{EBH} &= \sum_{i,\alpha} E_{\alpha} L_{\alpha \alpha}^i + ( - t_1 \sum _{ <i,j> } + t_2 \sum_{ <<i,j>> } )  \notag \\
& \quad \times \sum_{ \alpha \alpha^{\prime }\beta \beta^{\prime } } T_{ \alpha \alpha^{\prime }\beta \beta^{\prime }}^{ i j } L_{ \alpha \alpha^{\prime }}^i L_{ \beta \beta^{\prime }}^j,
\end{align}
where $E_{\alpha} $ is the mean-field energy per site, $L_{\alpha \alpha^{\prime }}^i \equiv | i \alpha \rangle \langle i \alpha^{\prime }|, T_{ \alpha \alpha^{\prime }\beta \beta^{\prime }}^{ i j } \equiv \langle i \alpha | \tilde{a} _i^{\dagger} | i \alpha^{\prime }\rangle \langle j \beta | \tilde{a}_j | j \beta^{\prime }\rangle $.

The single-particle retarded Green's function is defined as
\begin{align}
g_{i,j} (t-t^{\prime }) = - i \Theta(t-t^{\prime }) \langle [ a_i(t), a_i^{\dagger}(t^{\prime })] \rangle,
\end{align}
where $\Theta(t) $ is the step function. In the standard basis, the Green's function reads
\begin{equation}
g_{i,j}(t-t^{\prime }) = \sum_{ \alpha \alpha^{\prime }\beta \beta^{\prime
}}  T^{ji}_{\beta \beta^{\prime }\alpha \alpha^{\prime }} G^{i
j}_{ \alpha \alpha^{\prime }\beta \beta^{\prime }} (t-t^{\prime }),
\end{equation}
where
\begin{equation}
G^{i j}_{ \alpha \alpha^{\prime }\beta \beta^{\prime }} (t-t^{\prime }) = - i \Theta(t-t^{\prime }) \langle [ L^i_{\alpha \alpha^{\prime }}(t), L^j_{\beta \beta^{\prime }}(t^{\prime }) ] \rangle.
\end{equation}

By introducing the random-phase approximation, one obtains the equations of motion for $G$ in the frequency and momentum space,
\begin{align}  \label{equation of motion}
\delta_{\alpha \beta^{\prime }} \delta_{ \alpha^{\prime }\beta } D_{ \alpha \alpha^{\prime }} & = ( \omega - E_{\alpha^{\prime }} + E_{\alpha} ) G_{ \alpha \alpha^{\prime }\beta \beta^{\prime }}( k_x, \omega )  \notag \\
& \quad - D_{ \alpha \alpha^{\prime }} \sum_{ \gamma \gamma^{\prime }} \Big[  \epsilon( k_x+ k_{x c} ) \tilde{T}_{ \alpha^{\prime }\alpha \gamma \gamma^{\prime }}
\notag \\
& \quad + \epsilon( k_x - k_{x c} ) \tilde{T}_{ \gamma \gamma^{\prime }\alpha^{\prime}\alpha } \Big] G_{ \gamma \gamma^{\prime }\beta \beta^{\prime }} ( k_x, \omega ),
\end{align}
where $D_{\alpha \alpha^{\prime }} \equiv  \langle L_{\alpha \alpha} \rangle - \langle L_{ \alpha^{\prime}\alpha^{\prime }} \rangle $, $k_{x c} $ is given in Eq. (\ref{phi}) for superfluid phase and is zero for MI phase, $\tilde{T}_{ \alpha^{\prime }\alpha \gamma \gamma^{\prime }} \equiv y^{\dagger}_{\alpha^{\prime }\alpha} y_{\gamma \gamma^{\prime }}$, $y_{\alpha^{\prime }\alpha}^{\dagger} \equiv \langle i \alpha^{\prime }| e^{ i k_{xc} x_i} a_i^{\dagger} | i \alpha \rangle $, and $y_{\gamma \gamma^{\prime }} \equiv \langle i \gamma | e^{- i k_{xc} x_i} a_i | i \gamma^{\prime }\rangle $. $\tilde{T}_{\alpha^{\prime} \alpha \gamma \gamma^{\prime}}$ is site independent. Equations (\ref{equation of motion}) are linear equations of $\sum_{\alpha \alpha^{\prime }} y_{\alpha \alpha^{\prime }} G_{\alpha \alpha^{\prime }\beta \beta^{\prime }} $ and $\sum_{\alpha \alpha^{\prime }} y_{\alpha \alpha^{\prime }}^{\dagger} G_{\alpha \alpha^{\prime }\beta \beta^{\prime }}$. Substituting the solution into the Green's function $g(k_x,\omega)$, one obtains
\begin{equation}  \label{green function}
g( k_x, \omega ) = \frac{ \Pi ( k_x - 2 k_{x c}, \omega ) }{ 1 - \epsilon( k_x ) \Pi( k_x - 2 k_{x c}, \omega ) },
\end{equation}
where 
\begin{align}
\Pi (k_x, \omega) & = A_{11}( \omega) + \epsilon( k_x ) \frac{ A_{12} ( \omega) A_{21}(\omega) }{ 1 - \epsilon( k_x ) A_{22}( \omega ) }, 
\\
A_{11}(\omega) & = \sum_{\alpha} \left[ \frac{ y_{0 \alpha} y_{\alpha 0}^{\dagger} }{ \omega_{+} - \Delta E_{\alpha} } -\frac{ y_{ \alpha 0 } y_{ 0 \alpha}^{\dagger} }{ \omega_{+} + \Delta E_{\alpha} } \right], 
\\
A_{12}(\omega) & = A_{21}^{\dagger}(\omega)  
\notag \\
& = \sum_{\alpha} \left[ \frac{ y_{0 \alpha} y_{\alpha 0} }{ \omega_{+} - \Delta E_{\alpha} } -\frac{ y_{ \alpha 0 } y_{ 0 \alpha} }{ \omega_{+} + \Delta E_{\alpha} }\right], 
\\
A_{22}(\omega) & = \sum_{\alpha} \left[ \frac{ y_{0 \alpha}^{\dagger} y_{\alpha 0} }{ \omega_{+} - \Delta E_{\alpha} } -\frac{ y_{ \alpha 0 }^{\dagger} y_{ 0 \alpha} }{ \omega _{+} + \Delta E_{\alpha} } \right],
\end{align}
$| i \alpha=0 \rangle$ denotes the mean-field single-site ground state, $\omega_{+} = \omega + i 0^{+} $, and $\Delta E_{\alpha} = E_{\alpha} - E_0 $. Here the Green's function is a generalization of that in the standard BHM \cite{rpa}, in which bosons condense at zero momentum ($k_{x c}=0$).

\begin{figure}[htb]
\includegraphics[width=3.in]
{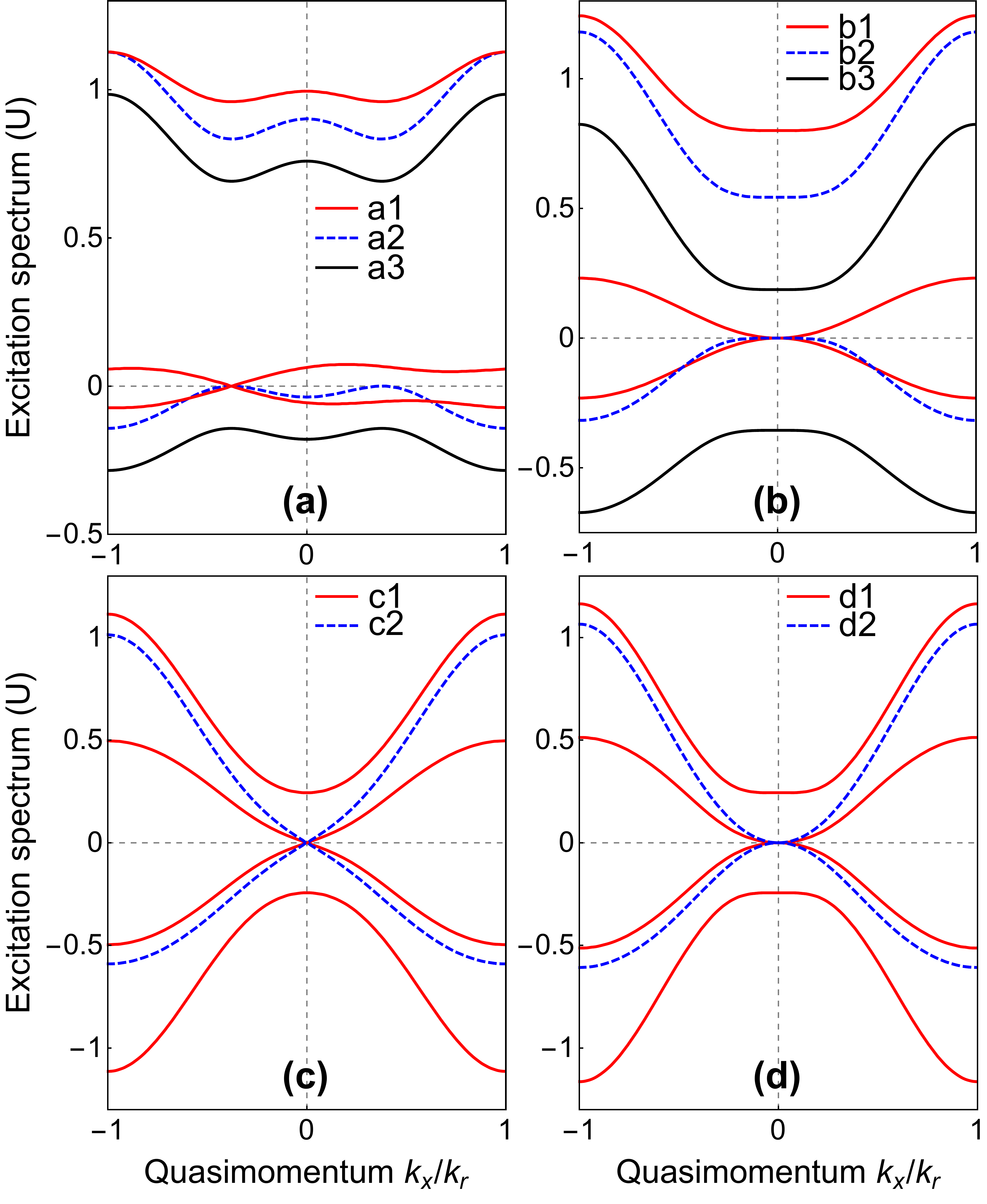}
\caption{Excitation spectra with parameters marked in Fig. \protect\ref{phase_diagram} (a) and (b). Red, blue (dashed), and black lines denote excitation spectra in the superfluid phase, at the MI-superfluid transition boundary, and  in the MI phase, respectively. }
\label{excitation}
\end{figure}

In the MI phase, the basis is just the Fock state $| i \alpha = n \rangle$. For a commensurate filling $n$, there is nonvanishing $y_{n^{\prime }n} = \sqrt{n} \delta_{n^{\prime }, n-1} $. The Green's function reads
\begin{equation}
g(k_x, \omega) = \frac{ Z }{ \omega_{+} - E_p } + \frac{ 1 - Z }{ \omega_{+} - E_h },
\end{equation}
where
\begin{align}
E_{p,h} & = \frac{1}{2} \Big[ U (2 n-1) - 2 \mu + \epsilon(k_x)
\notag \\
& \quad \pm \sqrt{ U^2 + 2 ( 2 n+1 ) U \epsilon(k_x) + \epsilon(k_x)^2 } \Big] 
\\
Z & = \frac{ \mu + U + E_p(k_x) }{ E_p(k_x) - E_h(k_x) }.
\end{align}
$E_{p,h}$ represents particle (hole) excitation. The condition of existence of a gapless excitation at $k_x= k_{x c}$ exactly gives the MI phase boundary in Eq. (\ref{MI boundary1}).

In the superfluid phase, we will numerically calculate the Green's function in Eq. (\ref{green function}) and the spectral function $\mathcal{A}(k_x,\omega)= - (1/\pi)$ Im $g(k_x,\omega)$, of which excitation modes give excitation spectra. Fig. \ref{excitation} shows excitation spectra near different MI-superfluid phase boundaries. There are two gapless spectra in the superfluid phase with positive and negative energy corresponding to quasiparticle and quasihole excitation, respectively.  In the superfluid there are also gapped excitation modes as a consequence of the band structure of the lattice system.  In the $Z_2$SF phase, roton excitation spectrum has been observed in the weakly-interacting regime \cite{roton}. Here we show the roton excitation spectrum in the strong-coupling limit in Fig. \ref{excitation} (a).  Figures \ref{excitation} (a) and (c) show linear dispersion around condensate momentum in the $Z_2$SF and NSF phase, respectively. Figures \ref{excitation} (b) and (d) show quadratic dispersions around $k_x=0$ at the NSF-$Z_2$SF transition boundary without and with particle-hole symmetry, respectively. In the superfluid phase, the quadratic dispersion indicates stronger phase fluctuations and weaker superfluidity than linear dispersion \cite{miao}. At the lower MI-superfluid transition boundary, the gapless particle dispersion vanishes, the hole dispersion becomes quadratic around $k_x = k_{xc}$, and a Mott gap is opened, which means disappearance of superfluidity. At the lobe tip, the Mott gap vanishes. In the Mott phase, the dispersions of both particle and hole excitations are gapped.

\section{conclusion}
In conclusion, we have shown a significant NNN hopping effect in near-resonantly shaken optical lattices. We studied the mean-field phase diagram for a one-dimensional EBH model and found tricritical points between three phases. Furthermore, we calculated corresponding microscopic parameters to the EBH model parameters and provided strong support for the existence of the tricritical point with particle-hole symmetry. A new critical behavior \cite{sf ising} is expected to be verified by the $\mathit{in}$ $\mathit{situ}$ technique \cite{in situ} in the parameters regimes. We also calculated the excitation spectra in all three phases and showed how the spectrum evolves during the phase transitions. In the $Z_{2}$SF phase, the excitation spectrum has the roton structure. At the NSF-$Z_{2}$SF transition boundary, the quasiparticle (quasihole) excitation has a quadratic dispersion relation around $k_{x}=0$. 

\section{acknowledgements}
We thank H. Zhai,  W. Zheng, C. Chin, and Y. Ohashi for valuable discussions and suggestions.


\begin{thebibliography}{99}
\bibitem{Hemmerich} L.-K. Lim, C. M. Smith, and A. Hemmerich, Phys. Rev. Lett. \textbf{100}, 130402 (2008).

\bibitem{Sengstock1} 
A. Eckardt, P. Hauke, P. Soltan-Panahi, C. Becker, K. Sengstock, and M. Lewenstein, Euro-phys. Lett. \textbf{89}, 10010 (2010).

\bibitem{Sengstock2} 
J. Struck, C. \"{O}lschl\"{a}ger, R. Le Targat, P. Soltan-Panahi, A. Eckardt, M. Lewenstein, P. Windpassinger, and K. Sengstock, Science \textbf{333}, 996 (2011).

\bibitem{Sengstock3}
J. Struck, C. \"{O}lschl\"{a}ger, M. Weinberg, P. Hauke, J. Simonet, A.
Eckardt, M. Lewenstein, K. Sengstock, and P. Windpassinger, Phys. Rev. Lett. \textbf{108}, 225304 (2012).

\bibitem{Tsuji} N. Tsuji, T. Oka, P. Werner, and H. Aoki, Phys. Rev. Lett. \textbf{106}, 236401 (2011).

\bibitem{haldane1} 
W. Zheng and H. Zhai, Phys. Rev. A \textbf{89}, 061603 (2014).

\bibitem{haldane2} 
G. Jotzu, M. Messer, R. Desbuquois, M. Lebrat, T. Uehlinger, D. Greif, and T. Esslinger, Nature (London) \textbf{515}, 237 (2014).

\bibitem{chin} 
C. V. Parker, L. C. Ha, and C. Chin, Nature Phys. \textbf{9}, 769 (2013).

\bibitem{sf ising} 
W. Zheng, B.-Y. Liu, J. Miao, C. Chin, and H. Zhai, Phys. Rev. Lett. \textbf{113}, 155303 (2014).

\bibitem{algebraic order1} 
H.-C. Po and Q. Zhou, arXiv:1408.6421 (2014).

\bibitem{miao} 
J. Miao, B. Liu, and W. Zheng, Phys. Rev. A \textbf{91}, 033404 (2015).

\bibitem{topo} 
S.-L. Zhang and Q. Zhou, Phys. Rev. A \textbf{90}, 051601(R) (2014).

\bibitem{bhm} 
M. P. A. Fisher, P. B. Weichman, G. Grinstein, and D. S. Fisher, Phys. Rev. B \textbf{40}, 546 (1989).

\bibitem{sf-mi exp} 
M. Greiner, O. Mandel, T. Esslinger, T. W. H\"{a}nsch, and I. Bloch, Nature \textbf{415}, 39 (2002).

\bibitem{t1t2} 
T. A. Zaleski and T. K. Kop\'{e}c, J. Phys. B: At. Mol. Opt. Phys. \textbf{43} 085303 (2010) .

\bibitem{t1t2_2}
 M. Di Liberto, O. Tieleman, V. Branchina, and C. M. Smith, Phys. Rev. A \textbf{84}, 013607 (2011).

\bibitem{nn int} 
G. Mazzarella, S. M. Giampaolo, and F. Illuminati, Phys. Rev. A \textbf{73}, 013625 (2006).

\bibitem{dipolar int} 
K. G\'{o}ral, L. Santos, and M. Lewenstein, Phys. Rev. Lett. \textbf{88}, 170406 (2002).

\bibitem{int hopping} 
T. Sowi\'{n}ski, O. Dutta, P. Hauke, L. Tagliacozzo, and M. Lewenstein, Phys. Rev. Lett. \textbf{108}, 115301 (2012).

\bibitem{spin} 
S. Tsuchiya, S. Kurihara, and T. Kimura, Phys. Rev. A \textbf{70}, 043628 (2004).

\bibitem{disorder1} 
T. Giamarchi and H. J. Schulz, Phys. Rev. B \textbf{37}, 325 (1988).

\bibitem{rashba} 
J.-H. Park, C. H. Kim, J.-W. Rhim and J. H. Han, Phys. Rev. B. \textbf{85}, 195401 (2012).

\bibitem{landau}
L. D. Landau and E. M. Lifshitz, \textit{Statistical Physics}, (Butterworth-Heinemann, Oxford, 1980).


\bibitem{in situ} 
X. Zhang, C.-L. Huang, S.-K. Tung and C. Chin, Science \textbf{335}, 1070 (2012).

\bibitem{standard basis} 
K. Sheshadri, H. Krishnamurthy, R. Pandit, and T. Ramakrishnan, Europhys. Lett., \textbf{22}, 257 (􏱗1993)􏱧.

\bibitem{rpa} 
Y. Ohashi, M. Kitaura, and H. Matsumoto, Phys. Rev. A, \textbf{73}, 033617 (2006).

\bibitem{roton}
L.-C. Ha, L. W. Clark, C. V. Parker, B. M. Anderson, and C. Chin, Phys. Rev. Lett. \textbf{114}, 055301 (2015).

\end{thebibliography}
\end{document}